\documentclass[iop, twocolappendix]{emulateapj}

\usepackage{color,ulem}
\usepackage{apjfonts}

\newcommand{\hb}{\ifmmode {\rm H\beta} \else H$\beta$ \fi}
\newcommand{\feii}{\ifmmode {\rm Fe\ II} \else Fe {\sc ii} \fi}
\newcommand{\oiii}{[O {\sc iii}]}
\newcommand{\obs}{\ifmmode {I_{\rm obs}} \else $I_{\rm obs}$ \fi}
\newcommand{\intri}{\ifmmode {I_{\rm int}} \else $I_{\rm int}$ \fi}
\newcommand{\broad}{\ifmmode {\xi} \else $\xi$ \fi}
\newcommand{\compare}{\ifmmode {{\cal S}} \else ${\cal S}$ \fi}
\newcommand{\template}{\ifmmode {{\cal T}} \else ${\cal T}$ \fi}
\newcommand{\poly}{\ifmmode {{\cal P}} \else ${\cal P}$ \fi}

\def\kms{$\rm km\,s^{-1}$}

\journalinfo{To appear in {\it The Astrophysical Journal}.}
\slugcomment{Received 2015 December 22;  accepted 2016 February 3}

\begin{document}

\title{Supermassive Black Holes with High Accretion Rates in Active Galactic Nuclei. \\
VI. Velocity-resolved Reverberation Mapping of H$\beta$ Line}

\author
{Pu Du\altaffilmark{1,*}, 
Kai-Xing Lu\altaffilmark{2,1}, 
Chen Hu\altaffilmark{1},
Jie Qiu\altaffilmark{1},
Yan-Rong Li\altaffilmark{1},
Ying-Ke Huang\altaffilmark{1}, 
Fang Wang\altaffilmark{6},
Jin-Ming Bai\altaffilmark{6}, 
Wei-Hao Bian\altaffilmark{8},
Ye-Fei Yuan\altaffilmark{9},
Luis C. Ho\altaffilmark{4,5} and
Jian-Min Wang\altaffilmark{1,3,*}\\
(SEAMBH Collaboration)}

\altaffiltext{1}
{Key Laboratory for Particle Astrophysics, Institute of High Energy Physics,
Chinese Academy of Sciences, 19B Yuquan Road, Beijing 100049, China}

\altaffiltext{2}
{Astronomy Department, Beijing Normal University, Beijing 100875, China}

\altaffiltext{3}
{National Astronomical Observatories of China, Chinese Academy of Sciences,
 20A Datun Road, Beijing 100020, China}

\altaffiltext{4}
{Kavli Institute for Astronomy and Astrophysics, Peking University, Beijing 100871, China} 

\altaffiltext{5}
{Department of Astronomy, School of Physics, Peking University, Beijing 100871, China} 

\altaffiltext{6}{Yunnan Observatories, Chinese Academy of Sciences, Kunming 650011, China}

\altaffiltext{8}{Physics Department, Nanjing Normal University, Nanjing 210097, China}

\altaffiltext{9}{Department of Astronomy, University of Science and Technology of China, Hefei 
230026, China}

\altaffiltext{*}{Corresponding author: (dupu,wangjm)@ihep.ac.cn}

\begin{abstract}
In the sixth of the series of papers reporting on a large reverberation mapping
(RM) campaign of active galactic nuclei (AGNs) with high accretion rates, we
present velocity-resolved time lags of \hb emission lines for nine objects
observed in the campaign during 2012$-$2013. In order to correct the
line-broadening caused by seeing and instruments before the analysis of
velocity-resolved RM, we adopt Richardson-Lucy deconvolution to reconstruct
their H$\beta$ profiles.  The validity and effectiveness of the deconvolution
are checked out by Monte Carlo simulation. Five among the nine objects show
clear dependence of time delay on velocity. Mrk 335 and Mrk 486 show
signatures of gas inflow whereas the clouds in the broad-line regions (BLRs) of
Mrk 142 and MCG +06-26-012 tend to be radial outflowing. Mrk 1044 is consistent
with the case of virialized motions.  The lags of the rest four are not
velocity-resolvable. The velocity-resolved RM of super-Eddington
accreting massive black holes (SEAMBHs) shows that  they have
diversity of the kinematics in their BLRs.  Comparing with the AGNs with
sub-Eddington accretion rates, we do not find significant differences in the
BLR kinematics of SEAMBHs.
\end{abstract}
\keywords{accretion, accretion disks -- galaxies: active -- quasars: supermassive black holes}

\section{Introduction}
Active galactic nuclei (AGNs) are very energetic sources powered by accretion
onto supermassive black holes (SMBHs) in centers of their host galaxies. In
optical/ultraviolet (UV) spectra of AGNs, the most prominent features are
plenty of broad emission lines whose full-width-half-maximum (FWHM) spans from
$\sim 10^3$ to $2\times 10^4$ \kms (e.g., \citealt{schmidt1963, osterbrock1986,
boroson1992, sulentic2000, shen2011} and references therein). These emission
lines originate from the gas in the broad-line regions (BLRs), which is 
photoionized by the continuum from central engines
\citep{osterbrock1989,ho2004}.  In the past decades, reverberation mapping (RM,
e.g., \citealt{bahcall1972, blandford1982, peterson1993}) has become the
mainstay in studying the kinematics and geometry of BLRs, as well as in
measuring the masses of SMBHs in the centers of AGNs.  It discerns the BLRs 
in time domain, instead of spatially, by  monitoring spectroscopically the
response of broad emission lines to the variations of continuum fluxes. RM
observation has been carried out for more than 50 AGNs by many different groups
\citep{peterson1993, peterson1998, kaspi2000, peterson2002, peterson2004,
    kaspi2007, bentz2008, bentz2009, denney2009, rafter2011, rafter2013,
barth2011, barth2013, barth2015, du2014, wang2014, du2015, shen2015}, and
measures successfully their time delays between the variations of continuum and
emission lines.  Enormous progress has been made in understanding the BLRs
through RM observations (see recent reviews, e.g., \citealt{gaskell2009,
popovic2012, netzer2013}).

However, a main objective of RM is to reconstruct the so-called velocity-delay
map (also known as ``transfer function"), which is a function of the time lag
and line-of-sight velocity of BLR gas, instead of only getting an averaged time
lag of the emission line \citep{blandford1982, horne1994, horne2004}. Due to
the requirements of homogeneous sampling, accurate calibration, excellent
signal-to-noise (S/N) ratios and high resolution for the reconstruction of
velocity-delay maps \citep{horne2004}, a relatively simpler analysis, that
measuring the time lags of emission line as a function of line-of-sight
velocity (velocity-resolved time lags), have been applied to about a dozen of
AGNs (e.g., \citealt{bentz2008, bentz2009, bentz2010, denney2009, denney2010,
grier2013}) and reveal the kinematics and geometry of their BLRs. It is very
similar to the projection of velocity-delay map on the velocity axis. Furthermore,
some more recent works have recovered the velocity-delay maps in several AGNs
successfully by maximum entropy methods \citep{bentz2010, grier2013} or Markov
Chain Monte Carlo method \citep{pancoast2012, pancoast2014b}. 

Since 2012, we have started a large RM campaign to monitor a sample of AGNs
with high accretion rates, especially super-Eddington accreting massive black
holes (SEAMBHs).  The goal of the campaign is to understand better the
super-Eddington accretion onto black holes, the BLR physics and their potential
as a new probe of cosmological distance \citep{wang2013, wang2014}. The
observations and light curves of \hb emission line in the first year
(2012-2013) have been published in \citeauthor{du2014} (2014, hereafter Paper
I) and \citeauthor{wang2014} (2014, hereafter Paper II), and the time lags of
their \feii emission are presented in \citeauthor{hu2015} (2015, hereafter
Paper III). The second stage of the campaign (2013-2014) have also been
published in \citeauthor{du2015} (2015, hereafter Paper IV). These samples
provide us opportunity to take a close look at the kinematics and geometry of
the BLRs in high accretion rate AGNs. In particular, the well-known BLR
radius-luminosity correlation ($R_{\rm BLR}-L_{5100}$ relationship, see
\citealt{kaspi2000, bentz2013}) is reported to show some dependence on
accretion rate (paper IV), implying that the BLRs are indeed different in AGNs
with high accretion rates from the normal ones. 
 
This paper presents the velocity-resolved time lags of the sources in the first
year of our SEAMBH campaign. Section \ref{sec:observation} gives a short review
of the observation and data reduction. Section \ref{sec:correct} describes in
detail the deconvolution method we used to correct line-broadening caused by
seeing and instruments, as well as a Monte Carlo simulation to evaluate the
effectiveness of the correction. In this section, we also show a comparison of
the \hb profiles before and after the deconvolution procedures. The obtained
velocity-resolved time lags are provided in Section \ref{sec:timelags}.
Section \ref{sec:discussion} gives a brief discussion and Section
\ref{sec:conclusion} a summary of the new results. 

\begin{figure}
\centering
\includegraphics[width=0.45\textwidth]{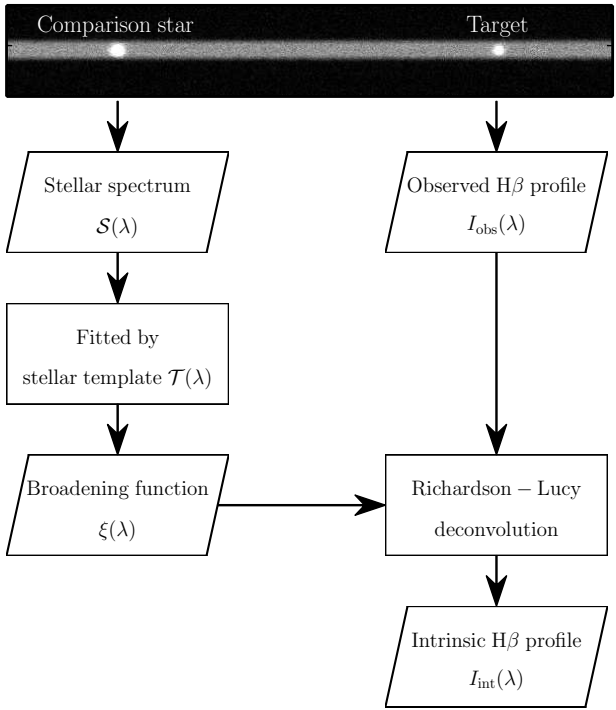}
\caption{Flow chart of correction for line-broadening.}\label{fig:flowchart}
\end{figure}

\section{Observations and Data Reduction}
\label{sec:observation}
Details of the target selection, the instruments, the observation and the
procedures of data reduction of the campaign in 2012-2013 (hereafter
SEAMBH2012) have been presented by Paper I and II. For completeness, in this
section, we provide a brief summary of the main points. 

\subsection{Sample Selection}
The targets selected for SEAMBH2012 are mainly narrow-line Seyfert 1 galaxies
(NLS1s, see, e.g., \citealt{osterbrock1985}), which, as a peculiar group of
AGNs, are suspected to be candidates of SEAMBHs. The characteristics of NLS1s
are: 1) narrow Balmer lines (FWHM $\lesssim2000$ km s$^{-1}$), 2) strong optical \feii
emission (relative to $\hb$), 3) weak \oiii\ lines and 4) steep hard X-ray
spectrum \citep{osterbrock1985, boller1996}. We selected NLS1s with extremely
steep 2$-$10 keV continuum, which is known to anti-correlate with accretion
rates (Wang et al. 2004).  For the SEAMBH2012 sample, we monitored nine NLS1s successfully,
whose names and coordinates are summarized in Table 1 of Paper II.

\subsection{Spectroscopy and Data Reduction}
All of the spectroscopic observations were carried out with the Yunnan Faint
Object Spectrograph and Camera which is mounted on the Lijiang 2.4-m telescope
at the Yunnan Observatories of the Chinese Academy of Sciences. To obtain spectra
with highly accurate flux calibration, we oriented the long slit to take the
spectra of the object and a nearby non-varying comparison star (used as
standard) simultaneously. This method was applied to RM campaign early in
\cite{maoz1990} and \cite{kaspi2000}. In SEAMBH2012, we fixed the slit width to
2\arcsec.5 given the seeing conditions, and used Grism 14 with a resolution of
92 \AA\ mm$^{-1}$ and a wavelength coverage of 3800-7200\AA\ which yields
$\sim1.8{\rm \AA\, pixel^{-1}}$ ($\sim108{\rm\ km\ s^{-1}\ pixel^{-1}}$) in
dispersion direction finally. All of the spectroscopic data were reduced with
IRAF v2.16 and the flux was calibrated by comparing spectra of the targets with
those of the comparison stars exposured in the slit.  Readers are referred
to Paper I for more details.

\begin{deluxetable}{lll}
\tabletypesize{\scriptsize}
\tablecaption{Stellar Template\label{tab:template}}
\tablewidth{0pt}
\tablehead{
\colhead{Object} & \colhead{Stellar Template} & \colhead{Spectral Type$^{\rm a}$}
}
\startdata
Mrk 335 & HD 77818 & K1 IV \\
Mrk 1044 & HD 173399 & G5 IV \\
IRAS 04416+1215 & HD 10307 & G1.5 V \\
Mrk 382 & SAO 126242 & F7 Vw \\
Mrk 142 & HD 142373 & F8 Ve... \\
MCG +06-26-012 & HD 131156 & G8 V \\
IRAS F12397+3333 & HD 118100 & K5 Ve \\
Mrk 486 & HD 20618 & G6 IV \\
Mrk 493 & HD 118100 & K5 Ve
\enddata
\tablecomments{Name and spectral type of the selected template for the comparison stars.}
\tablenotetext{a}{The spectra type listed in \cite{valdes2004}.}
\end{deluxetable}

\section{Correction for Line-broadening}
\label{sec:correct}
In spectroscopic observation, the line-broadening ($\broad$), which is caused
by seeing and instruments (e.g., grism or grating), influences the observed
profiles of emission lines.  Although the instrumental settings remain the same
during observations, the seeing varies nights by nights (FWHM of the
point-spread functions in SEAMBH2012 is generally from 1\arcsec.2 to
2\arcsec.5, while the width of slit is fixed to 2\arcsec.5 which is wider than
seeing), and \broad in each night changes as well. Furthermore, the
mis-centering of objects and comparison stars in the slit is not a constant in
different nights. This causes that each spectrum have slightly different shift
in dispersion direction.  Fortunately, we are also able to estimate the shifts
from the spectra of comparison stars so that we can correct this automatically
by deconvolution.  

The observed H$\beta$ profile of $\obs$ is a convolution of its intrinsic
profile $\intri$ and the line-broadening function $\broad$, namely 
\begin{equation}
\obs(\lambda) = \intri(\lambda) \otimes \broad(\lambda) = \int \intri(\lambda^{\prime}) 
                 \broad(\lambda - \lambda^{\prime}) d\lambda^{\prime}.
\end{equation}
Variations of \broad does not change the integrated flux of \hb emission line,
however, it blurs \intri and further influences the velocity-resolved time
series analysis. It is thus expected to correct the line-broadening of H$\beta$
profiles before the velocity-resolved RM analysis, especially for those objects with relatively
narrow H$\beta$ lines (FWHM $\sim$ 1000 -- 2000km/s, comparing to the
instrumental broadening of $\sim$500km/s estimated in Paper I) in our
SEAMBH2012 observation. Fortunately, the spectra of comparison stars in the
same slit during spectroscopic exposures can be employed to estimate $\broad$ accurately 
in light of that the FWHM of absorption lines in stellar spectra ($\lesssim$ 10\kms) is much
narrower than the line-broadening. An efficient way to perform such a
correction is the well-known Richardson-Lucy algorithm \citep{richardson1972,
lucy1974} which can  deconvolve \broad from \obs so as to recover $\intri$. 

Figure \ref{fig:flowchart} shows the sketch of procedures for line-broadening
correction.  We first estimate the broadening function \broad of exposure via
fitting the spectrum of comparison star in each night by a stellar template,
then use the obtained \broad to correct the broadening of \hb profile \obs in
the observed spectra through Richardson-Lucy deconvolution algorithm. Finally,
we can get the broadeing-corrected profile \intri. 

\begin{figure}
\centering
\includegraphics[width=0.45\textwidth]{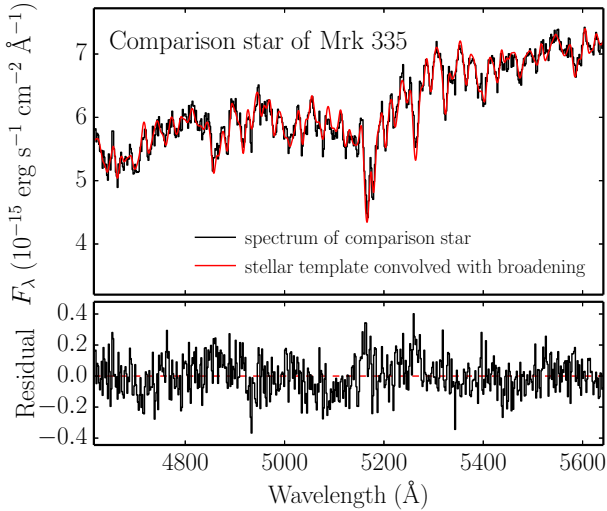}
\caption{An individual-night spectrum of the comparison star of Mrk 335 and its best fitting model. The top panel shows 
the observed spectrum (black line) of the star and the best model (red line). The bottom panel is the residual.}\label{fig:fitting}
\end{figure}

\subsection{Richardson-Lucy Deconvolution}
Richardson-Lucy (R-L) deconvolution algorithm is a Bayesian-based iterative
technique commonly used for recovering signal blurred by a known response. It
has been widely adopted to improve the spatial resolution of images from Hubble
Space Telescope ({\it HST}, e.g., \citealt{lauer1993, snyder1993, lauer1995,
gebhardt1996, schmitt1996, kormendy1999, farrah2001, rest2001, lauer2005} and
the references therein), and to correct the instrumental broadening in
spectroscopic data such as ultraviolet spectra from Goddard High Resolution
Spectrograph on {\it HST} (e.g., \citealt{wahlgren1991, brandt1993,
bomans1996}) or X-ray spectroscopy measurements of heavy elements
\citep{fister2007}. Even recently, \cite{menezes2014} used it in data analysis
of integral field spectrograph from Gemini. To our knowledge, this is the first
application of such technique to the correction of instrumental broadening in
spectra from RM observation.

We do not repeat the details of R-L deconvolution here, but just give a brief
description (readers may refer to original papers \cite{richardson1972} and
\cite{lucy1974}). In short, if $I_{\rm int}^n$ is the $n$-th estimate of the
intrinsic profile in the iterative sequence, the $(n+1)$-th estimate is
obtained by
\begin{equation}
I_{\rm int}^{n+1}(\lambda^{\prime}) = I_{\rm int}^n(\lambda^{\prime}) \int \frac{I_{\rm obs}(\lambda)}{I_{\rm obs}^n(\lambda)} 
        \broad(\lambda - \lambda^{\prime}) d\lambda,
\end{equation}
where
\begin{equation}
I_{\rm obs}^n(\lambda) = \int I_{\rm int}^n(\lambda^{\prime}) \broad(\lambda - \lambda^{\prime}) d\lambda^{\prime}.
\end{equation}
Similar to the widely used equidistribution of ignorance in Bayes's postulate,
we assume a uniform distribution as the initial guess of intrinsic profile, so
that 
$I_{\rm int}^0(\lambda)=\int \obs(\lambda) d\lambda \ / \int d\lambda$. 
After a certain number (see Section \ref{sec:termination}) iterations, the
intrinsic profile can be recovered effectively. 

\begin{figure}
\centering
\includegraphics[width=0.45\textwidth]{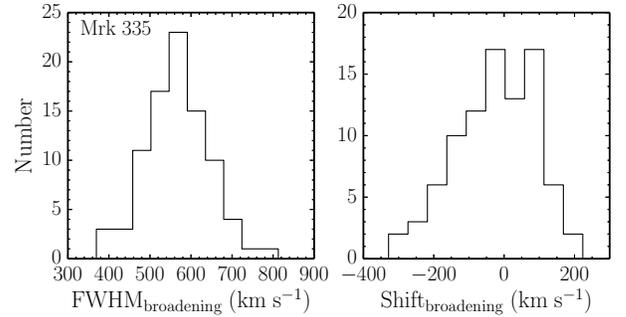}
\caption{The distribution of the broadening functions \broad of Mrk 335
obtained by fitting the stellar spectra.  The left panel shows the distribution
of FWHM of \broad in each night, and the right panel shows the distribution of their
shifts.}\label{fig:mrk335_broad}
\end{figure}

\subsection{Line-broadening Function}
\label{sec:broad}
During the spectroscopic exposures, we oriented the slit in order to
simultaneously include the target and a nearby comparison star for their
spectra (an object-star pair). It enables us to obtain \broad of the object
synchronously from the comparison star in the pair. To estimate the broadening
function, we use the direct pixel-fitting method similar to \cite{greene2006}
and \cite{ho2009}.  The observed spectrum \compare of comparison star is
modeled by a stellar template spectrum \template convolved by the broadening
function: 
\begin{equation}
\compare(\lambda) = \poly(\lambda) \left[\template(\lambda) \otimes \broad(\lambda) \right].
\end{equation}
Here, \poly is a polynomial accounted for large-scale mismatch in the continuum
shape of the comparison star and template. In this work, we assume \broad a
Gaussian function and select the most similar stellar spectrum as the template
of comparison star from the Indo-U.S. Library of Coud\'{e} Feed Stellar
Spectra\footnote{http://www.noao.edu/cflib/} \citep{valdes2004}. The selected
templates for the comparison stars are listed in Table \ref{tab:template}. In
order to get the line-broadening function close enough to $\hb$, we set the
fitting window to 4500\AA-5500\AA\ in rest frame of the objects. Prior to the
fitting, the consecutive exposures of the object-star pair in each night are
combined so as to provide high S/N ratio. The best values of
FWHM and shift in \broad are determined by fitting the combined spectrum of the
comparison star in each night via Levenberg-Marquardt algorithm. 

An example of individual-night spectrum of the comparison star and the best
model for Mrk 335 is shown in Figure \ref{fig:fitting}. The distributions of
FWHM and shift of \broad during the observation of Mrk 335 are shown in Figure
\ref{fig:mrk335_broad}. The FWHM, on average, is consistent with the mean value
of 500km s$^{-1}$ estimated in Paper I. However, through the fitting here, we
are able to measure the broadening function in each individual night. For the
other objects, the distributions of the broadening are similar to the case of
Mrk 335, so we do not show them repeatedly here.  It should be noted that,
although SEAMBHs generally have weak \oiii$\lambda5007$, we can still try to
estimate \broad from the \oiii\ line in several objects with relatively strong
and intrinsicly narrow \oiii\ (e.g., Mrk 382) by comparing its width to the
value of spectra with higher resolution \citep{whittle1992}, if we assume the
narrow lines do not vary with time.  The \broad obtained from \oiii\ is
consistent with the values estimated by the fitting here, and the differences
are $\lesssim50$km s$^{-1}$ averagely. 

After getting \broad in each night (for each target), we reconstruct \intri by
applying R-L deconvolution to the observed spectra \obs of the objects.

\begin{figure}
\centering
\includegraphics[width=0.48\textwidth]{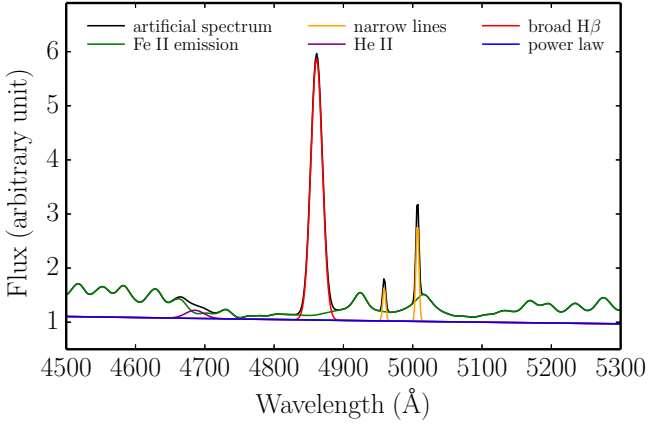}
\caption{Artificial AGN spectrum in the Monte Carlo simulation. The black line is the created artificial AGN spectrum. The components are: 
(a) a power-law continuum (blue); (b) \feii emission (green) and broad He {\sc ii} (purple); 
(c) narrow emission lines of \oiii$\lambda\lambda$4959,5007 (orange) and (d) a broad component of \hb (red). } \label{fig:input}
\end{figure}

\subsection{Iteration Number}
\label{sec:termination}
The termination of iterations is one of the core issues in R-L 
deconvolution. Too few iterations are insufficient for reconstruction of the
original signal, while an excess of iterations magnifies high-frequency noise
and even produces nonsense results \citep{lucy1974}. Here, a simple Monte Carlo
approach is used to determine the appropriate number for terminating the
iterations. 

For each object, we first construct an Gaussian, with the same FWHM and
equivalent width (EW) of \hb in the observed mean spectra, where the mean
broadening of 500 km s$^{-1}$ estimated in Paper I have been subtracted, as the
artificial intrinsic spectrum. This artificial intrinsic spectrum is used to
generate a mock observed spectrum \obs in each individual night by convolving
with \broad obtained from the fitting in Section \ref{sec:broad} and adding
random noises with the same level of S/N as the truly observed spectrum. Then,
we select the iteration number which can get the best reconstruction (with the
smallest $\chi^2$) of the artificial intrinsic spectrum by de-convolving \broad
from the mock \obs as the optimal number of iterations. The mock \obs is
generated 20 times for each spectrum, and the median value of the optimal
numbers of iterations obtained by the above procedure is chosen as the final
iteration number of the deconvolution.

\subsection{Host Galaxy Subtraction}
The spectra of AGNs and comparison stars are smeared by the same broadening
function every night since both of them are point sources. However, the host
galaxies of the objects are extended and their spectra are affected by
broadening which is different from those of comparison stars. Therefore, before
de-convolving the spectra of AGNs with the broadening function obtained from
the comparison stars in Section \ref{sec:broad}, the contribution of host
galaxies should be removed. Paper III has already decomposed the spectra and
extracted the contribution of host galaxies by fitting scheme. We simply
subtract those host contributions, which we got in Paper III, from the spectra
of objects before the procedures of deconvolution to avoid potential influence
of different broadening in host galaxies.

\begin{figure}
\centering
\includegraphics[width=0.48\textwidth]{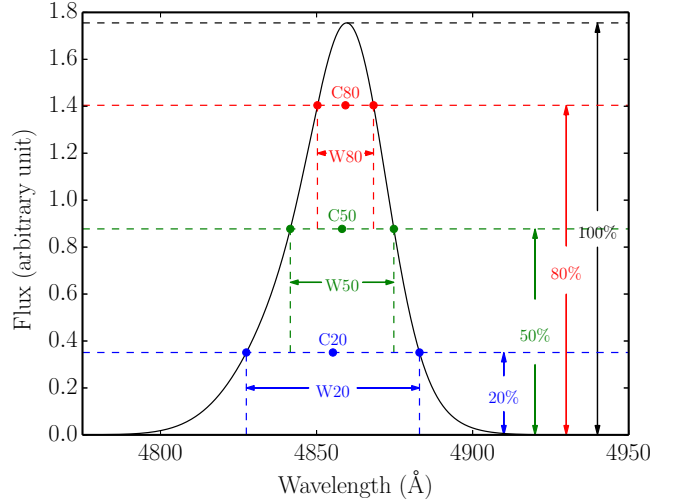}
\caption{Definition of (W20, W50, W80) and (C20, C50, C80). These parameters can be used as indicators to 
characterize the profile of emission line.} \label{fig:W_C}
\end{figure}

\begin{figure*}
\centering
\includegraphics[width=0.8\textwidth]{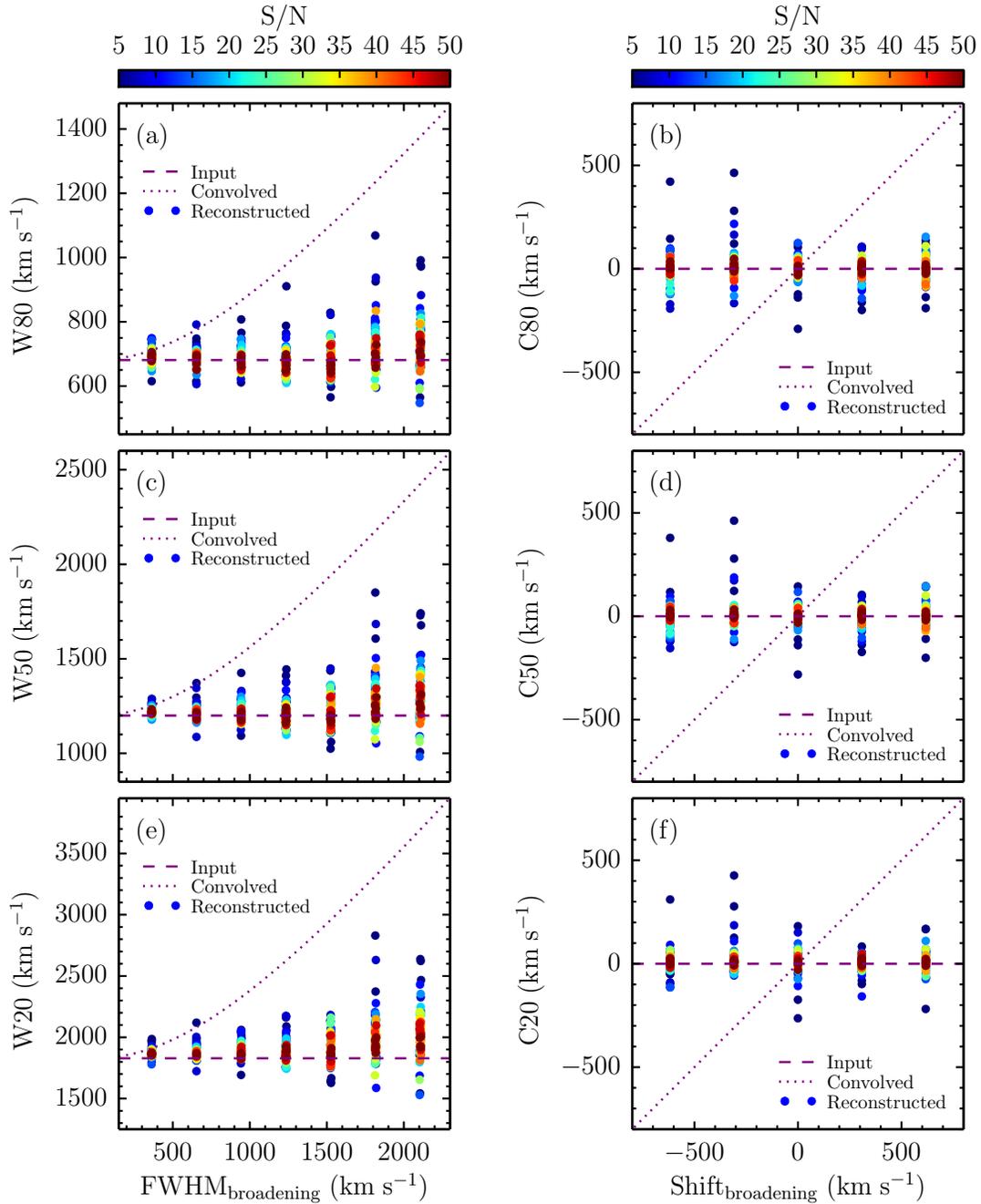}
\caption{Comparisons of \hb profiles between the input and output spectra in
the Monte Carlo simulation.  FWHM$_{\rm broadening}$ and Shift$_{\rm
broadening}$ are the width and shift of broadening function \broad. The dashed
lines mark the values of (W20, W50, W80) and (C20, C50, C80) in the input
artificial spectrum, while the dotted lines show their values in the convolved
output spectra. The points are (W20, W50, W80) and (C20, C50, C80) of the
profiles in the reconstructed spectra after the deconvolution. The colors show
the S/N ratios we set in the simulated observed spectra. The reconstruction is
more successful in the cases of narrower broadening function and higher S/N
ratios.}
\label{fig:fwhm}
\end{figure*}

\subsection{Monte Carlo Simulation on Deconvolution}
\label{sec:mc}
We make use of Monte Carlo simulation to evaluate the robustness of
reconstructing the intrinsic profile of \hb line through the R-L deconvolution.
We first create an artificial AGN spectrum by including (a) a power-law
continuum; (b) \feii emission and broad He {\sc ii}; (c) narrow emission lines
of \hb and \oiii$\lambda\lambda$4959,5007 and (d) a broad component of $\hb$.
The components here are almost the same as the model we used in Paper III
except for the absence of host galaxy contribution (host does not influence the
profiles of \hb in the simulation).  We set the FWHM and EW of \hb in the
artificial AGN spectrum to 1200km/s and 100\AA\ (see Figure \ref{fig:input}),
respectively, which are the typical values in the present sample. In order to
test the precision of recovery by the deconvolution in the cases with various
broadening and S/N ratio, we broaden the artificial spectrum by different
broadening function and plus stochastic noise to generate the simulated
observed spectra. In the meantime, we select a stellar spectral template (HD
44951 used here) from the Indo-U.S. Library and broaden it with the same
broadening as the AGN spectrum and plus noise as well. We take the simulated
observed AGN and star spectra as the input of the procedures in Section
\ref{sec:correct}, and then compare the profiles of \hb in the output
reconstructed spectra to those of the original artificial AGN spectra.

To elaborate the comparison of profiles between the spectra before and after the
    deconvolution procedures, we define full width at (20\%, 50\%, 80\%) of
    maximum as (W20, W50, W80), and centers at (20\%, 50\%, 80\%) of maximum as
    (C20, C50, C80) to characterize the profiles of \hb (please see Figure \ref{fig:W_C} for
    their definition). (W20, W50, W80) evaluate the width of emission line at different height, 
    while (C20, C50, C80) show the shifts at the corresponding levels (see figure \ref{fig:W_C}). 
    Here, W50 is just simply FWHM in the common sense. The comparison of emission-line
    profiles before and after the deconvolution in the Monte Carlo simulations
is shown in Figure \ref{fig:fwhm}.  Generally speaking, the deconvolution can recover the profile of \hb correctly. 
(W20, W50, W80) and (C20, C50, C80) of the reconstructed \hb profiles are consistent with the
value of the input artificial AGN spectrum in the cases with high S/N ratios.  
The reconstruction is more successful 
in the cases with narrower broadening function and higher S/N ratio, especially when the FWHM of broadening 
is narrower than the input \hb profile. The widths of \hb in the observed spectra of SEAMBH2012 are 
1000 $\sim$ 2000 km s$^{-1}$, while the broadening is only $\sim$ 500 km s$^{-1}$. And the S/N ratio of spectra in 
SEAMBH2012 is generally higher than 20. Therefore, the intrinsic profiles 
recovered by the deconvolution Section \ref{sec:correct} is reliable.

\subsection{Profiles Before and After the Deconvolution}
\label{sec:comparison}
\begin{figure}
\centering
\includegraphics[width=0.48\textwidth]{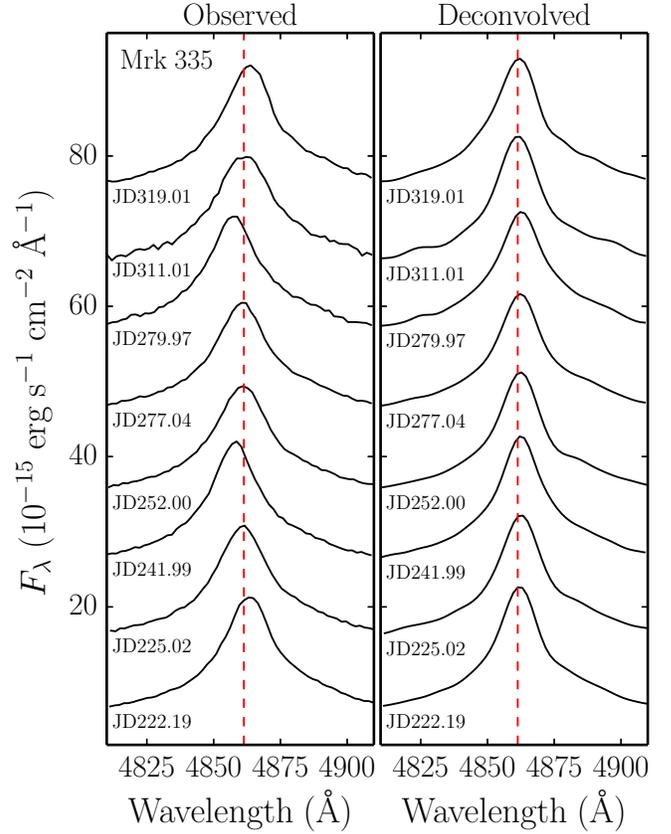}
\caption{A comparison of deconvolved profiles (Mrk 335) with the observed ones. The profiles are shifted 
vertically by a constant value ($10^{-14}\ {\rm erg\ s^{-1}\ cm^{-2} \AA^{-1}}$) from bottom to 
top for recognition. The broadening caused by instruments and varying seeing is corrected successfully. 
The mis-centering effect (see Section \ref{sec:correct}) is also eliminated effectively. The dashed lines 
mark the center of $\hb\lambda4861$. The wavelength is in the rest-frame and JD-2456000 is marked next to 
each profile.} 
\label{fig:profiles}
\end{figure}

\begin{figure*}
\centering
\includegraphics[width=0.95\textwidth]{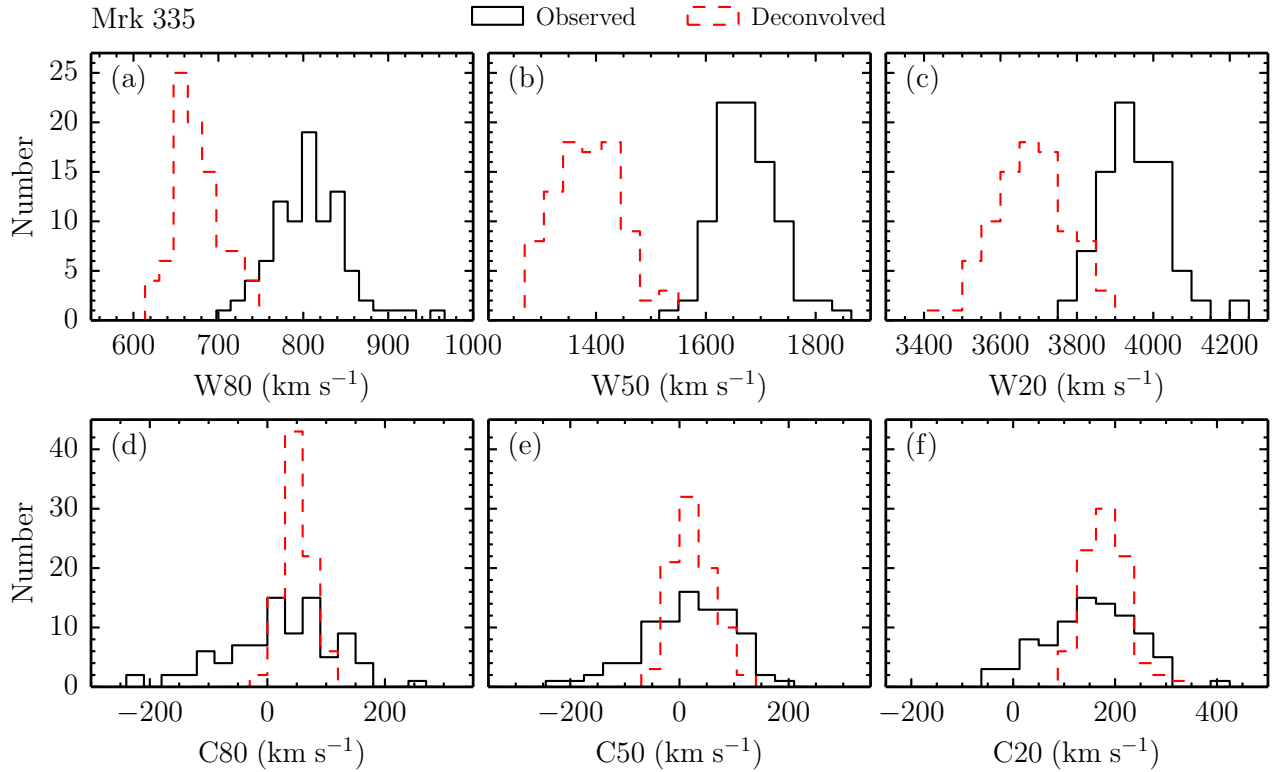}
\caption{A comparison of widths and shifts before and after the deconvolution
(Mrk 335). Panels (a) - (f) are W80, W50, W20, C80, C50 and C20, respectively. 
The black solid lines are the distributions in the observed
profiles, and the red dashed lines are the distributions in the reconstructed
profiles after the deconvolution.}
\label{fig:width_shift}
\end{figure*}

In order to give an overall evaluation how effectively does the deconvolution
work, examples of the comparison between observed and de-convolved spectra of
Mrk 335 are provided in Figure \ref{fig:profiles}. It is very obvious that the
deconvolution procedures correct the broadening and mis-centering effect
successfully. To further compare the profiles before and after the
deconvolution, we measure (W20, W50, W80) and (C20, C50, C80) of \hb in the
spectra before and after the deconvolution procedures. As an example, the
values of Mrk 335 are shown in Figure \ref{fig:width_shift}. After the
deconvolution, the widths (W20, W50, W80) of its \hb lines narrow down for
$\sim$($6\%$, $15\%$, $20\%$) on average. In particular, the distribution of
W80 in the reconstructed spectra becomes much narrower, which means the more
serious influence caused by the broadening around the peak (the highest and
narrowest part) of emission line, than in its wing part (with broader width),
have been corrected effectively.  Additionally, the centers at different levels
(C20, C50, C80) are more concentrated than the cases before the deconvolution.
The comparison before and after the deconvolution demonstrates the validity of
the deconvolving procedures used in the present paper. The centers, it should
be noted, do not locate on 0 km s$^{-1}$ because the zero here is defined by the
redshift queried from NASA/IPAC Extragalactic
Database\footnote{http://ned.ipac.caltech.edu/}.

\section{Velocity-resolved Time Series Analysis}
\label{sec:timelags}
\subsection{Methods}
The light curves and the time delays between the variations of the continuum at
5100\AA\ and \hb fluxes measured by simply integrating the full extent of the
emission lines have been published in Paper I and II. Paper III updated these
measurements by introducing a sophisticated scheme that fit AGN continuum, host
contribution in the slit and emission lines simultaneously. The updated time
lags are consistent with the previous ones but have smaller uncertainties (see
the comparison in Paper III). However, these time delays only stand for the
radii averaged by the emissivity function of BLRs, and no information of
velocity fields is carried out. In this section, we focus on the
velocity-resolved analysis of delays in order to reveal the kinematics of 
BLR clouds in SEAMBHs.

The procedures of deconvolution in Section \ref{sec:correct} have removed the
varying line-broadening from the observed profiles and almost corrected the
wavelength shifts caused by mis-centering. After further correcting the
remanent shifts ($\lesssim 50$ km s$^{-1}$) using \oiii$\rm \lambda5007$ line
as our wavelength reference, the spectra are ready for velocity-resolved time
series analysis. RMS (root of mean-square) spectra of the objects are obtained
in the standard way from the spectra after the line-broadening correction.
Similar to many previous works (e.g., \citealt{bentz2008, bentz2009,
denney2009, denney2009b, bentz2010, denney2010, grier2013}), we divide \hb
emission line into several bins in velocity space, where each bin has equal
amount of flux in the RMS spectrum. After integrating the flux of emission line
in each bin, we cross-correlate the obtained individual-bin light curves to the
continuum light curve at 5100\AA.  Due to the lack of error bars in the output
spectra of deconvolution, we estimate the uncertainties of flux in the
individual-bin light curves based on the differences between adjacent points
using median filter as in Paper I. The interpolation cross-correlation function
(ICCF; \citealt{gaskell1986, gaskell1987, white1994}) is employed, and the
centroid of the CCF, above 80\% of the maximum cross-correlation coefficient
($r_{\rm max}$), is adopted as the time lags (e.g., \citealt{koratkar1991,
peterson2004, denney2006, bentz2009, denney2010} and reference therein).  The
uncertainties in the measured lag time are determined through the Monte Carlo
``flux randomization/random subset sampling" method described in
\cite{maoz1989} and \cite{peterson1998, peterson2004}. The rest-frame
velocity-resolved time lags of the objects, as well as their RMS spectra
created after deconvolution, are shown in Figure \ref{fig:mrk335_compare} and
\ref{fig:2d}. 
The details of objects are discussed below.

\subsection{Individual Objects}
\begin{figure*}
\centering
\includegraphics[width=0.42\textwidth]{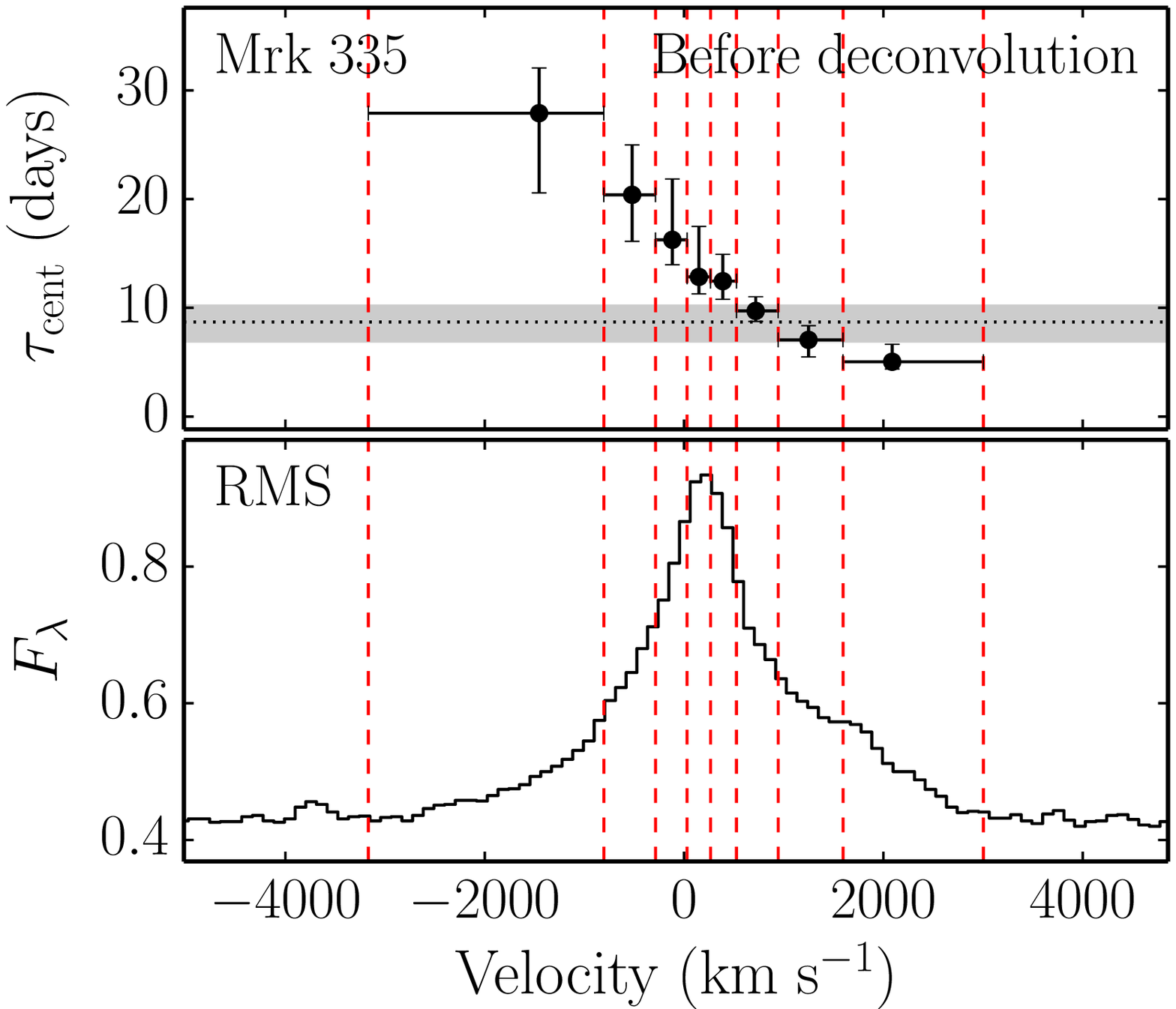} 
\hspace{0.02\textwidth}
\includegraphics[width=0.42\textwidth]{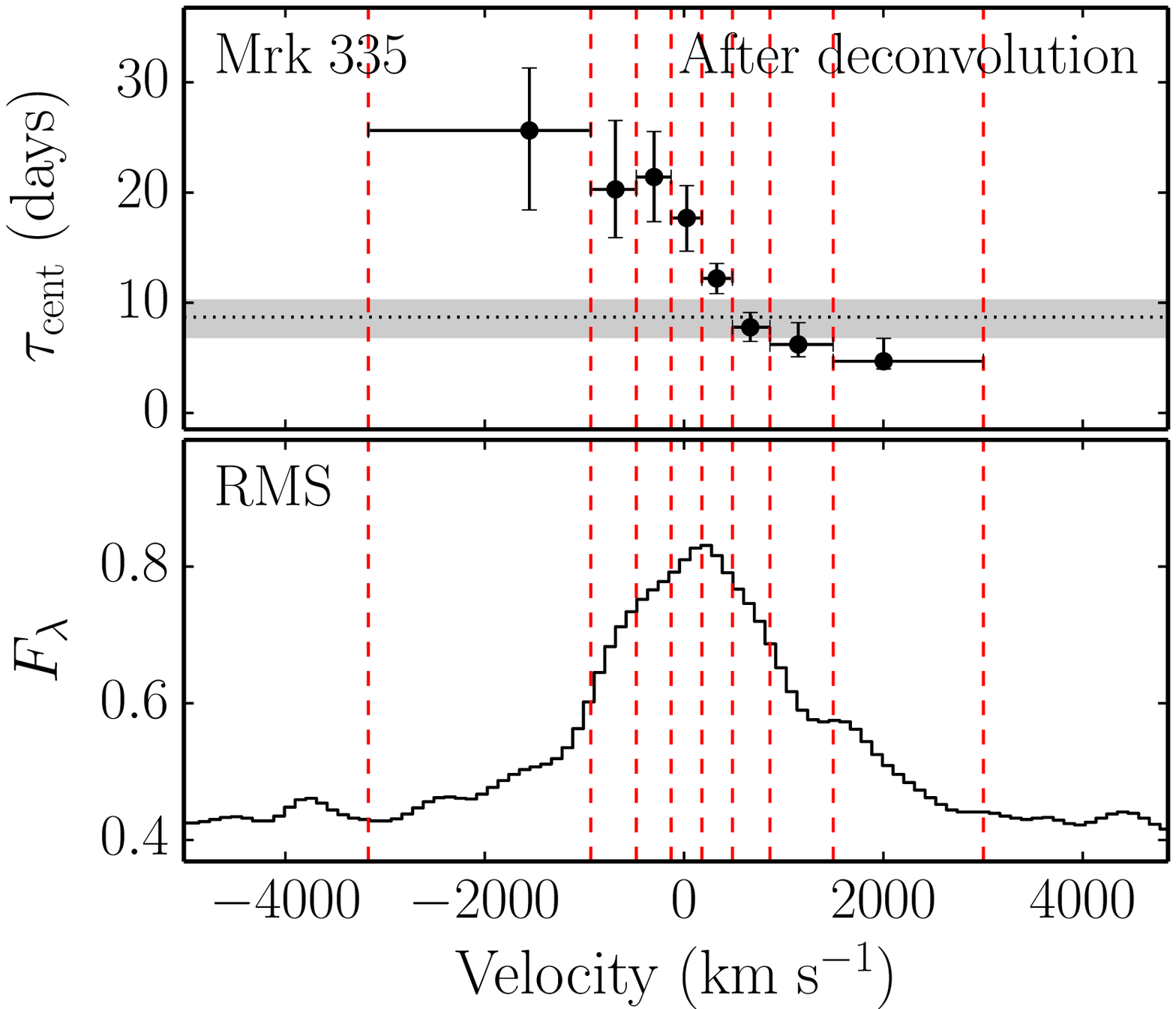}
\caption{Rest-frame velocity-resolved time delays and the corresponding RMS
spectra.  The result of Mrk 335 obtained from the spectra without deconvolution
is shown in the left plot, and its result from deconvolved spectra is shown in
the right one.  The upper panel of each plot shows the centroid time lags in
the divided velocity bins, and the lower panel is their RMS spectrum (in units
of ${\rm 10^{-15}\ erg~s^{-1}~cm^{-2}~\AA^{-1}}$).  The red dashed lines are
edges of the bins. The horizontal dotted line and the gray band mark the
average time lag and the uncertainties provided in Paper III. The horizontal
error bar in upper panel denote the velocity bin, while the vertical error bar
is the uncertainty of the lag.}\label{fig:mrk335_compare}
\end{figure*}

\textbf{Mrk 335.} The \hb emission line of this object is divided into eight
bins in velocity space, where each bin contains $1/8$ of the variable line flux
in its RMS spectrum. Its velocity-resolved time series analysis (see Figure
\ref{fig:mrk335_compare}) demonstrates clear gradient in the delays of
different velocity bins that the blueshifted part of \hb lags the response in
its redshifted part (from $\sim25$ days in the blue end to $\sim5$ days in the
red end).  This kinematic signature is consistent with the expectation of
inflow that the gas on near side is receding to the observer while the gas on
opposite side is approaching.  In Figure \ref{fig:mrk335_compare}, we also show
the result before the deconvolution. Generally speaking, the velocity-resolved
time delays before and after the deconvolution are consistent with each other.
However, the deconvolution-corrected result show some details, more clearly,
that the longer time delays in blueshifted part vary gently while the lags in
the redshifted part decrease more quickly. The before-deconvolution time lags
only show uniform gradient from blue- to red-shifted velocity. In particular,
the RMS spectrum before the deconvolution contains relatively stronger
contribution of narrow \hb emission line which is caused by the varying width
of narrow line (because of the changing broadening in each night), while this
feature is much weaker in the RMS spectrum of the reconstructed profiles. The
comparison in Figure \ref{fig:mrk335_compare} also indicates that the
deconvolution procedures can recover the velocity-resolved information
moderately in RM time-series analysis. 

Before our observation, Mrk 335 was monitored twice in 1989$-$1996
\citep{kassebaum1997, peterson1998} and 2010$-$2011 \citep{grier2012},
respectively. Its velocity-resoved time lags in the latter campaign have been
published in \cite{grier2013}.  It is not unexpected that the inflow signature
in the velocity-resoved delays presented here is consistent with the previous
measurements in 2010$-$2011 \citep{grier2013}, given the short time interval
between these two campaigns. Such a kind of signature has also been reported in
the velocity-resolved delay analysis of many other sources like Arp 151
\citep{bentz2008, bentz2009}, NGC 3516 \citep{denney2009, denney2010}, Mrk 1501
\citep{grier2013} and PG 2130+099 \citep{grier2013}. 

\textbf{Mrk 142.} The velocity-resolved time delays of Mrk 142 are reported
here for the first time.  It was previously monitored by the LAMP collaboration
\citep{bentz2009}, however the authors do not show its velocity-resolved lags
perhaps due to their data quality. The observation in our campaign has more
homogeneous cadence and better calibration precision. Opposite to the case of
Mrk 335, Mrk 142 shows signature of outflow indicated by the shorter time lag
from BLR gas in the blueshifted side of \hb line comparing with the gas in
redshifted side. Interestingly, to our knowledge, the signature of outflow was
observed distinctly by reverberation mapping in only one case, NGC 3227
\citep{denney2009, denney2010}, before the detection shown here. Comparing with
the low accretion rate in NGC 3227 (\citealt{denney2010}; Paper IV), Mrk 142
is a SEAMBH and radiation pressure acting on the ionized gas is probably the
driver for the observed outflow.

\textbf{Mrk 486.} It also demonstrates clear signature of infalling gas as the
case of Mrk 335. Except for the first bin, the response to the continuum
variation in the blueshifted gas lags the response of the gas corresponding to
the redshifted side. 

\textbf{Mrk 1044.} The velocity-resolved lags of Mrk 1044 are relatively
symmetric and show evidence of simple virialized motions, that high-velocity
gas is located in the more central region, which is similar to the cases of NGC
5548 \citep{denney2009}, SBS 1116+583A \citep{bentz2009}, Mrk 1310
\citep{bentz2009}, NGC 4051 \citep{denney2009b, denney2010} and 3C 120
\citep{grier2013} in the previous works. 

\textbf{MCG +06-26-012.} A signature of outflow is shown in this object that
the time delays ($\sim20$ days) in the blue part of \hb are slightly shorter
than those ($\sim30$ days) in the red part, however, the outflow signs are
ambiguous a little bit because of the large errors. The large error bars are
caused by the inhomogeneous sampling and the short monitoring period comparing with 
the variation in the light curves. MCG +06-26-012 is the only sub-Eddington
object in the present sample (see details in Paper II).

\textbf{Mrk 382.} The lags in different velocity bins are undistinguished. This
phenomenon has been found in other objects, such as Mrk 1310 \citep{bentz2009},
NGC5548 \citep{bentz2009} and Mrk 290 \citep{denney2010}.  This may be either
caused by that the lag differences among the bins are smaller than their error bars
(by the quality of the current data), or it could be real intrinsically that
those bins have the same time lags.  If the line-of-sight velocity keeps a constant in the
BLR, motion of cloud are fully chaotic. This remains open for future
observations with high quality data.

\textbf{IRAS F12397+3333} and \textbf{Mrk 493} are similar to Mrk 382.

\textbf{IRAS 04416+1215.} The direct integrating method in Paper I and II is
failed to give the average time lag of its integrated \hb emission line flux.
Only multi-components fitting method in Paper III is able to detect its delay.
Here, we adopt the direct integrating method to measure the light curves after
the division of bins and still fail to get reliable detection of its 
velocity-resolved lags. We divide its \hb into two bins and demonstrate the
result here for completeness. The $r_{\rm max}$ in the redshifted bin is only
$\sim0.4$.

\begin{figure*}
\centering
\includegraphics[width=0.42\textwidth]{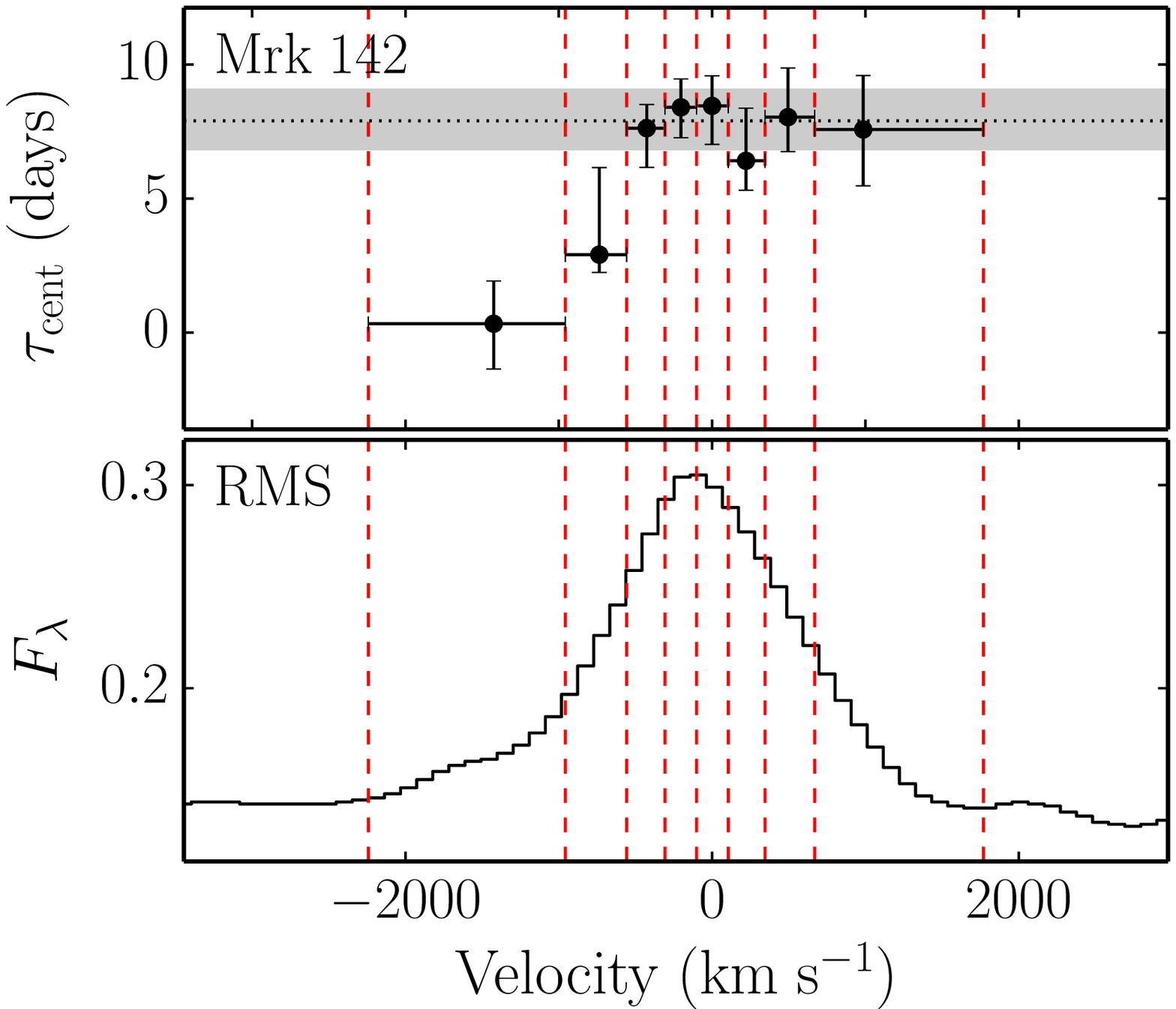}
\hspace{0.02\textwidth}
\includegraphics[width=0.42\textwidth]{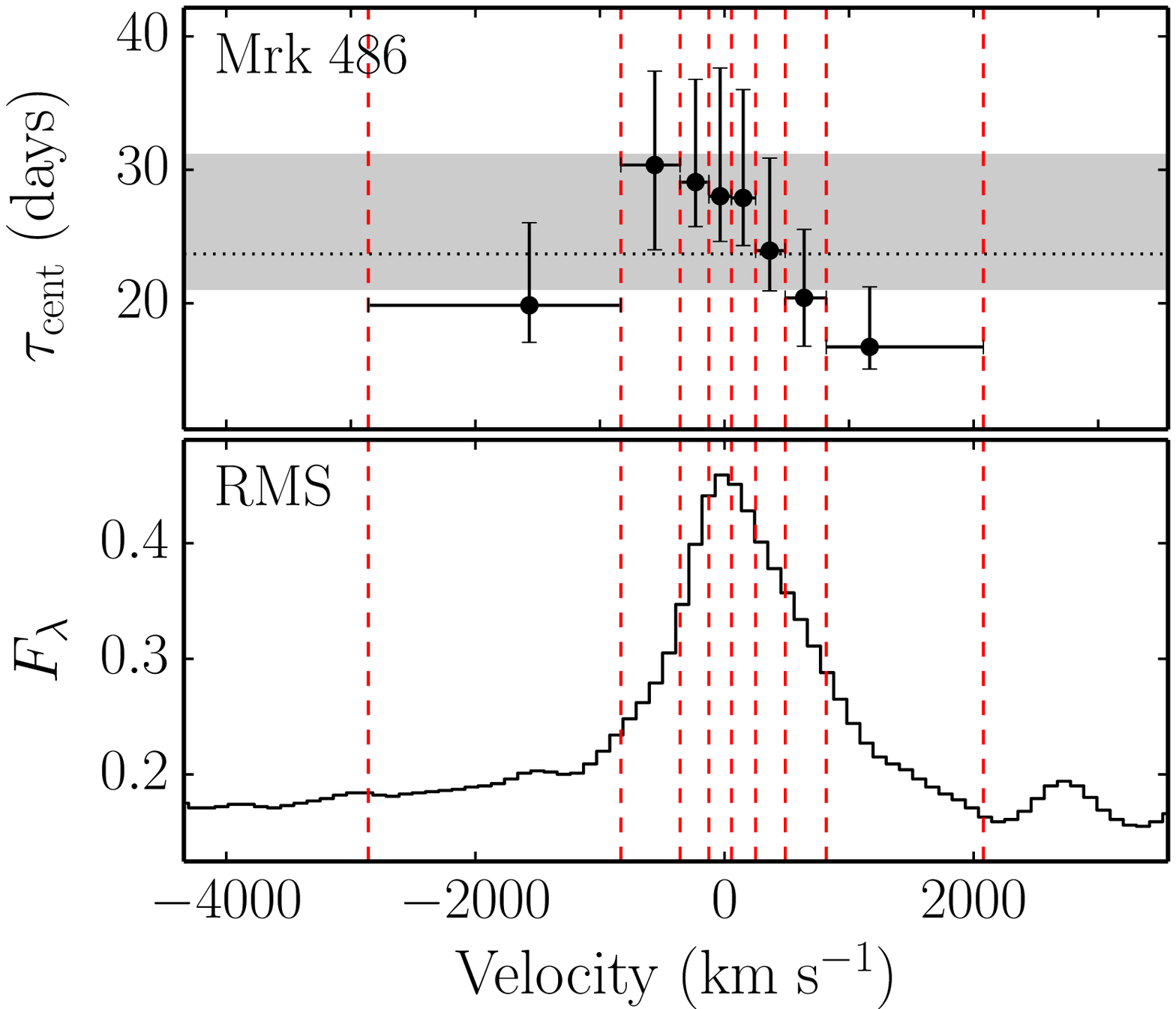} \\
\vspace{0.01\textwidth}
\includegraphics[width=0.42\textwidth]{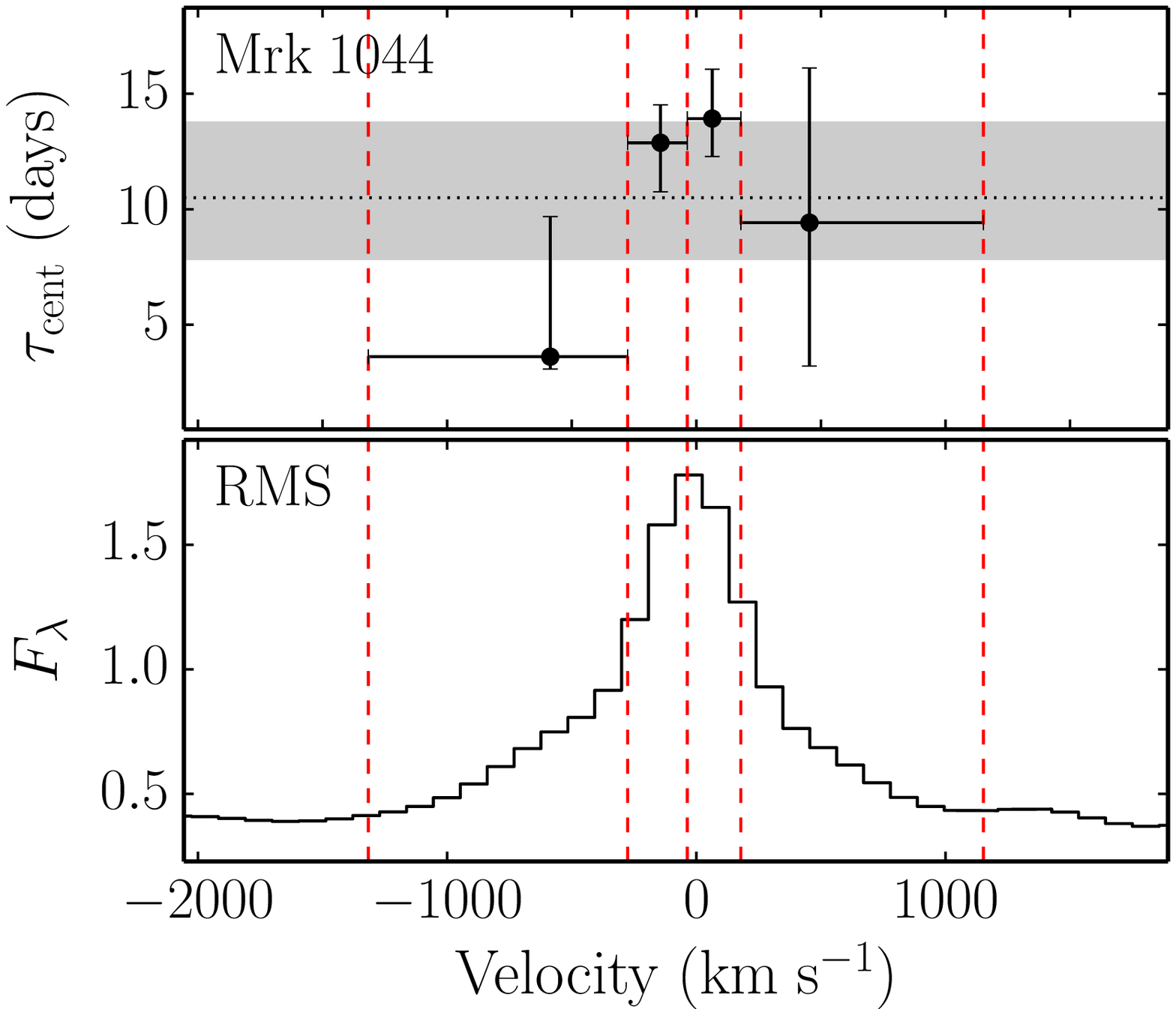}
\hspace{0.02\textwidth}
\includegraphics[width=0.42\textwidth]{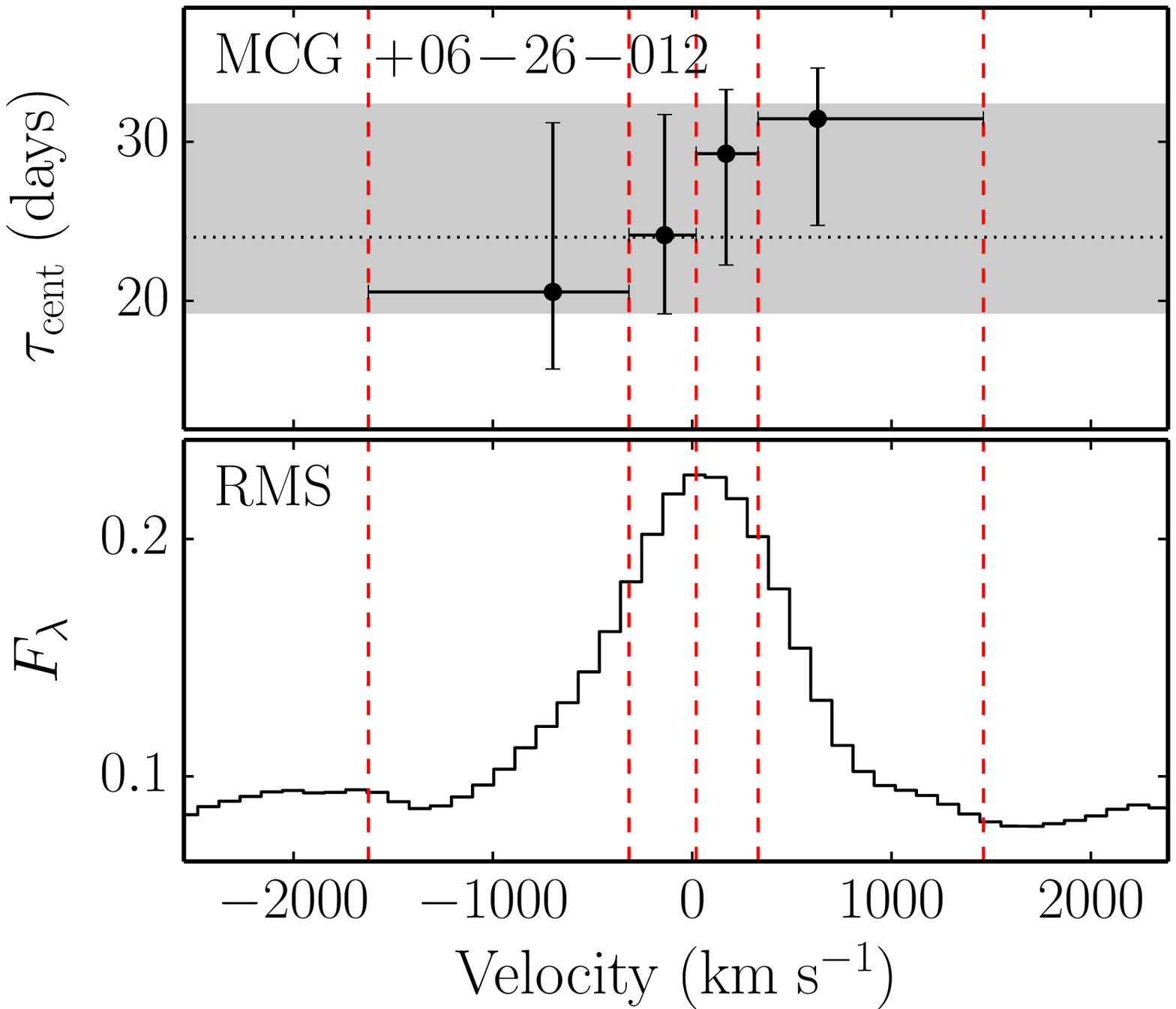}
\caption{Rest-frame velocity-resolved time delays and the corresponding RMS
spectra. The meanings of panels, symbols, lines and units are the same as in
Figure \ref{fig:mrk335_compare}. }\label{fig:2d}
\end{figure*}

\begin{figure*}
\figurenum{\ref{fig:2d}}
\centering
\includegraphics[width=0.42\textwidth]{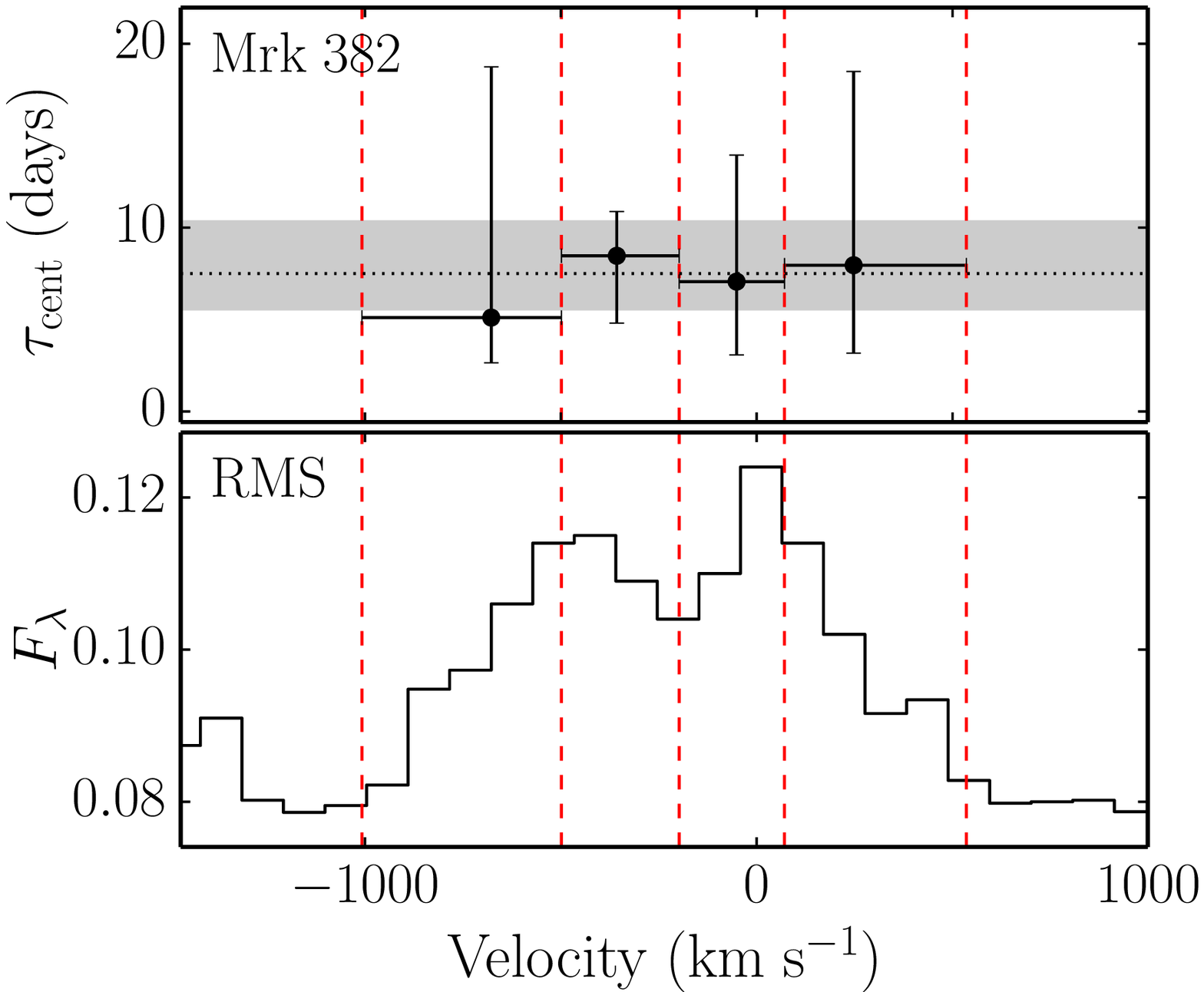}
\hspace{0.02\textwidth}
\includegraphics[width=0.42\textwidth]{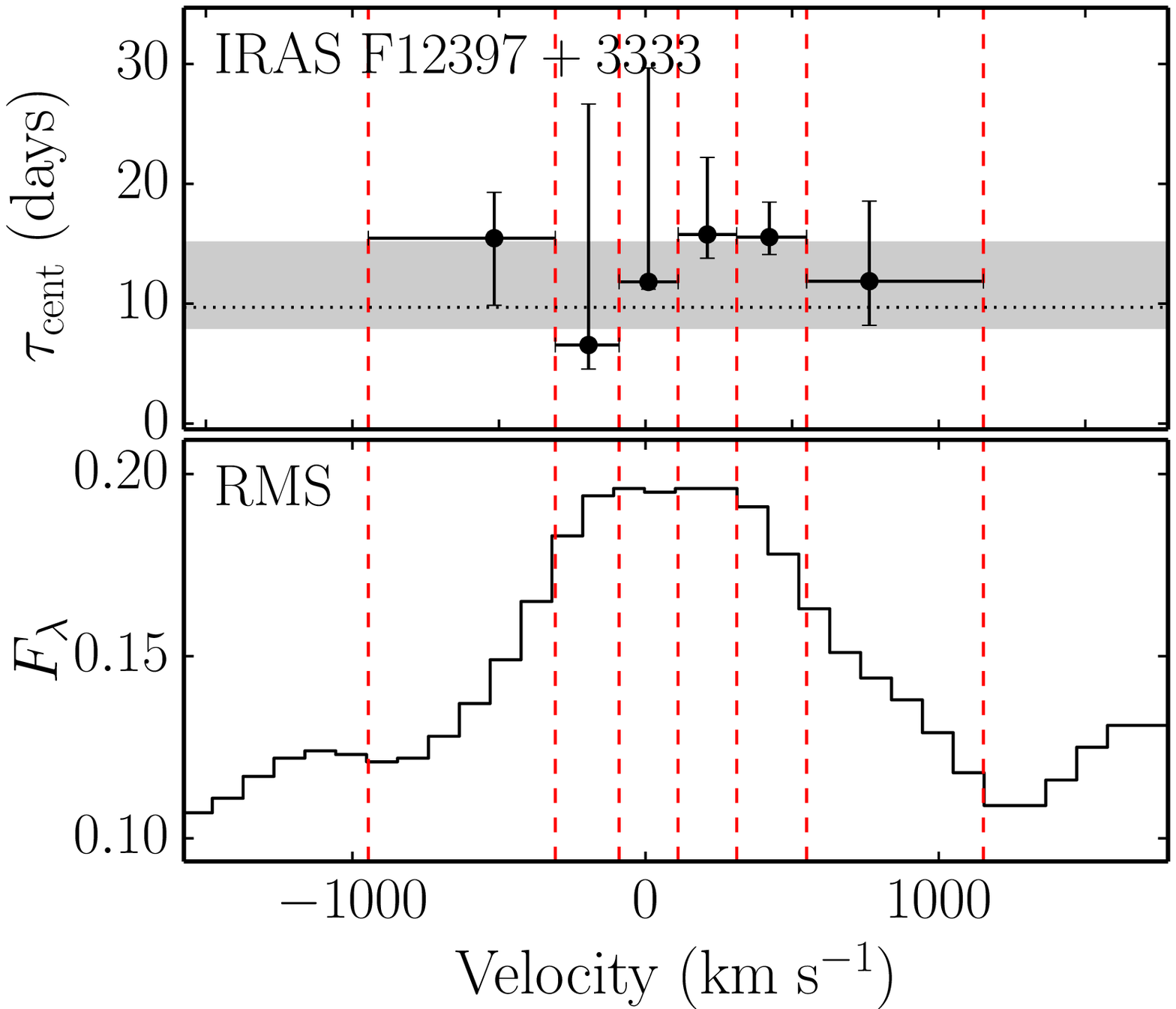} \\
\vspace{0.01\textwidth}
\includegraphics[width=0.42\textwidth]{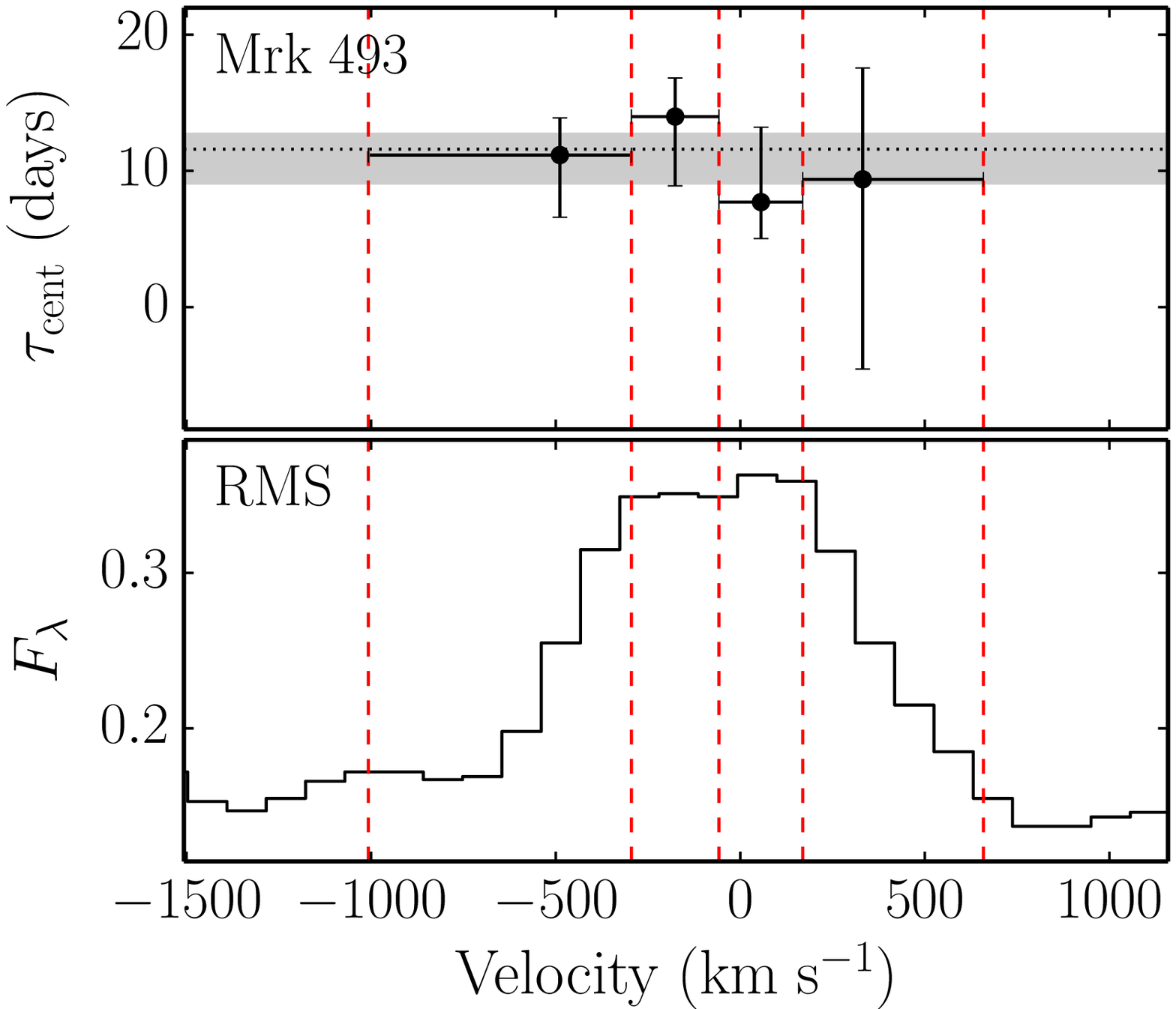}
\hspace{0.02\textwidth}
\includegraphics[width=0.42\textwidth]{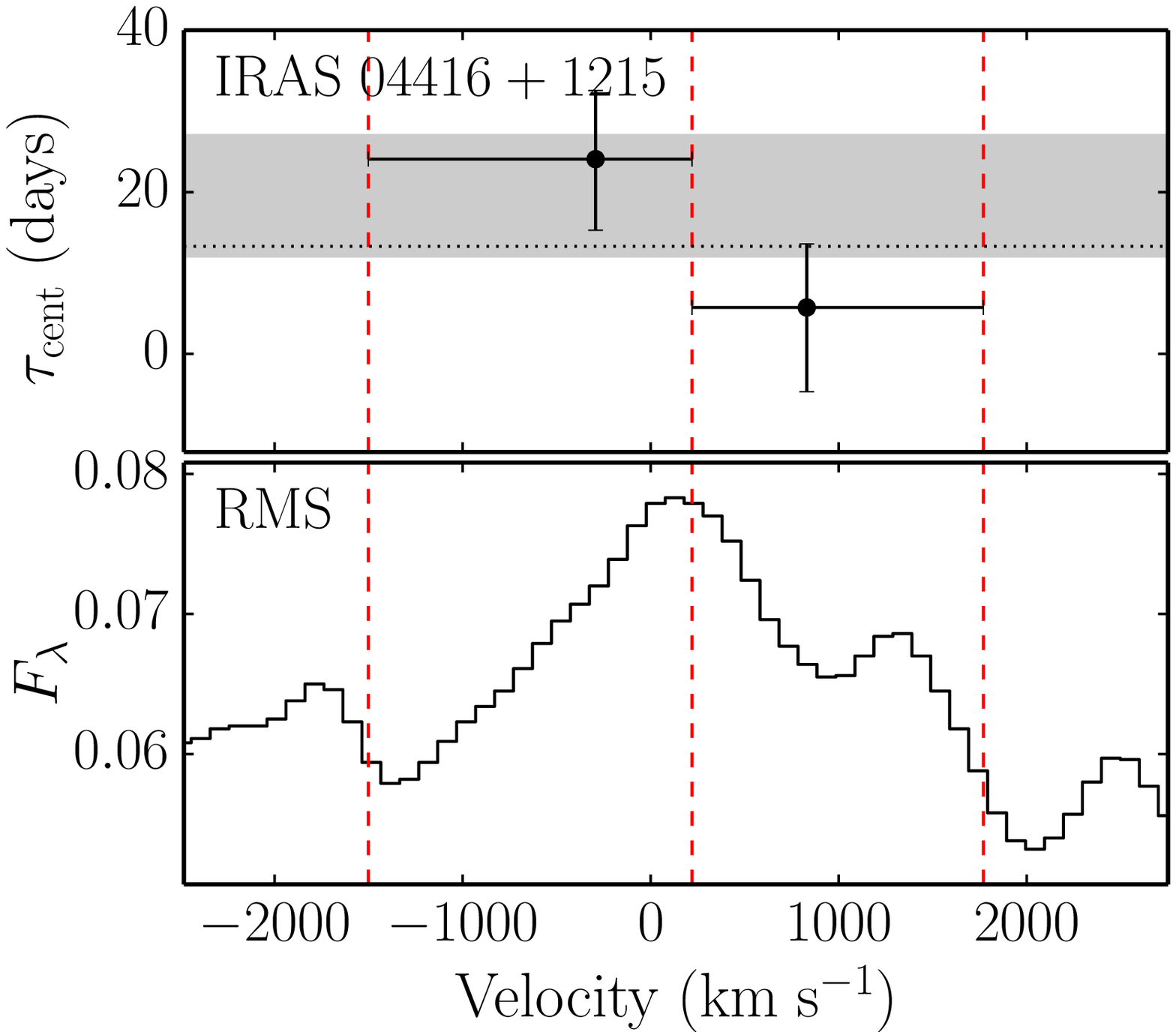}
\caption{Continued.}
\end{figure*}

\section{Discussions}
\label{sec:discussion}
%
Outflows suggested by the observation of blueshifted broad absorption lines are
not rare in AGNs (e.g., \citealt{trump2006, gibson2009} and reference therein).
However, how do the outflows contribute to broad emission lines, especially in
high accretion rate AGNs, is still subject to some debate.  Considering that
super-Eddington accreting AGNs probably have the strongest radiation pressure,
which can provide kinetic energy to BLR gas, among the whole AGN populations,
the possibilities of outflows that occur in SEAMBHs maybe doubt to be higher
than sources with normal accretion rates. However, the present sample, although
with its limited size, shows diverse properties of the BLR kinematics, rather
than the simple outflows (there is only one from the detected four SEAMBHs,
except for MCG +06-26-012 with low accretion rate). It is consistent with the
fact that the observed profile of \hb in the present sample does not show
unified signature of blueshifted asymmetry.  On the other hand, there are a few
objects (2 out of 4 super-Eddington AGNs) showing clear evidence for the
presence of infalling gas in BLRs. Considering the fact that objects in the
present sample are super-Eddington accreting AGNs, the velocity fields of the
surrounding gas showing infall in kinematics are not surprising.  Otherwise, a
significant fraction of the accreting gas will be channelled into outflows,
making fast growth of less massive black holes be suppressed. It is impossible
to reach a conclusion about the pattern of velocity fields of ionised gas in
the vicinity of black holes in super-Eddington accreting AGNs. We hope
increases of monitored SEAMBHs and normal AGNs could clarify the patterns of
the velocity fields.


\section{Conclusions}
\label{sec:conclusion}
We have presented here velocity-resolved time lag measurements of nine objects
observed in our SEAMBH2012 campaign. To correct the line-broadening caused by
instruments and seeing, we use Richardson-Lucy deconvolution to reconstruct \hb
profiles. The broadening function in each night is estimated by fitting the
spectra of the stars observed simultaneously with stellar templates, and Monte
Carlo method is adopted to test the validity of the broadening-correction
procedures. We show that the Richdson-Lucy iteration is an efficient method to
recover the velocity-resolved information to some extent. 
    
All of the objects in the present sample are NLS1s with very high accretion
rates except for MCG +06-26-012. Five sources among them demonstrate clear
velocity-resolved time delays in their \hb emission lines. The
velocity-resolved lags of Mrk 335 and Mrk 486 show the signature of gas infall,
while the BLR clouds in Mrk 142 and MCG +06-26-012 tend to be radial
outflowing. The symmetric pattern of Mrk 1044 is consistent with virialized
motions. The time lags of the other four objects are not velocity-resolvable.
We do not find significant differences in the BLR kinematics of SEAMBHs,
comparing with sub-Eddington AGNs.  Some discussions for the result of each
individual object are provided in the main text. This analysis provides new
insight into the geometry and kinematics of BLRs in SEAMBHs.

\acknowledgements{We acknowledge the support of the staff of the Lijiang 2.4m
telescope. Funding for the telescope has been provided by CAS and the People's
Government of Yunnan Province. This research is supported by the Strategic
Priority Research Program - The Emergence of Cosmological Structures of the
Chinese Academy of Sciences, Grant No. XDB09000000, by NSFC grants
NSFC-11173023, -11133006, -11373024, -11233003, -11503026, -11573026 and -11473002, 
a NSFC-CAS joint key grant U1431228, the China-Israel nsfc-isf 11361140347 and 
by the Key Research Program of the Chinese Academy of Sciences, Grant No. KJZD-EW-M06.}

\end{document}